\documentclass[preprintnumbers,amsmath,amssymbm,preprint]{revtex4}
\usepackage{epsfig}
\usepackage{graphicx}

\begin{document}
\title{Nontrivial spatial behavior of the Gauss-Bonnet curvature invariant of rapidly-rotating Kerr black holes}
\author{Shahar Hod}
\affiliation{The Ruppin Academic Center, Emeq Hefer 40250, Israel}
\affiliation{ }
\affiliation{The Hadassah Institute, Jerusalem 91010, Israel}
\date{\today}

\begin{abstract}
\ \ \ The Gauss-Bonnet curvature invariant has attracted the attention of physicists and mathematicians 
over the years. In particular, it has recently been proved that black holes can 
support external matter configurations that are non-minimally coupled to the 
Gauss-Bonnet invariant of the curved spacetime. 
Motivated by this physically interesting behavior of black holes in Einstein-Gauss-Bonnet theories, 
we present a detailed {\it analytical} study of the physical and mathematical properties of the 
Gauss-Bonnet curvature invariant ${\cal G}_{\text{Kerr}}(r,\cos\theta;a/M)$ 
of spinning Kerr black holes in the spacetime region outside the horizon 
[here $\{r,\theta\}$ are respectively the radial and polar coordinates of 
the black-hole spacetime, and $a/M$ is the dimensionless angular momentum of the black hole]. 
Interestingly, we prove that, for all 
spinning Kerr spacetimes in the physically allowed regime $a/M\in[0,1]$, 
the spin-dependent maximum curvature of the Gauss-Bonnet invariant is attained at the equator of the black-hole surface. 
Intriguingly, we reveal that the location of the global minimum of the Gauss-Bonnet invariant has a highly 
non-trivial functional dependence on the black-hole rotation parameter: (i) For Kerr black holes 
in the dimensionless slow-rotation $a/M<(a/M)^{-}_{\text{crit}}=1/2$ regime, 
the Gauss-Bonnet curvature invariant attains its global minimum asymptotically 
at spatial infinity, (ii) for black holes in the intermediate spin regime 
$1/2=(a/M)^{-}_{\text{crit}}\leq a/M\leq(a/M)^{+}_{\text{crit}}=
\sqrt{\Big\{{{7+\sqrt{7}\cos\Big[3^{-1}\arctan\big(3\sqrt{3}\big)\Big]-
\sqrt{21}\sin\Big[3^{-1}\arctan\big(3\sqrt{3}\big)\Big]\Big\}}/12}}$, the global minima are located at the black-hole poles, and 
(iii) Kerr black holes in the super-critical (rapidly-spinning) regime $a/M>(a/M)^{+}_{\text{crit}}$ 
are characterized by a non-trivial (non-monotonic) functional behavior of the Gauss-Bonnet 
curvature invariant ${\cal G}_{\text{Kerr}}(r=r_+,\cos\theta;a/M)$ along the black-hole horizon with a spin-dependent 
polar angle for the global minimum point. 
\end{abstract}
\bigskip
\maketitle

\section{Introduction}

Astrophysically realistic black holes rotate about their polar axis and it is therefore 
widely believed that they belong to the two-dimensional family \cite{Notetwo} of spinning Kerr spacetimes which 
are described by the curved line element \cite{ThWe,Chan,Notebl,Noteun}
\begin{eqnarray}\label{Eq1}
ds^2=-{{\Delta}\over{\rho^2}}(dt-a\sin^2\theta
d\phi)^2+{{\rho^2}\over{\Delta}}dr^2+\rho^2
d\theta^2+{{\sin^2\theta}\over{\rho^2}}\big[a
dt-(r^2+a^2)d\phi\big]^2\  .
\end{eqnarray}
The metric functions in (\ref{Eq1}) are given by the functional expressions
\begin{equation}\label{Eq2}
\Delta\equiv r^2-2Mr+a^2\ \ \ \ ; \ \ \ \ \rho^2\equiv r^2+a^2\cos^2\theta\  .
\end{equation}
The physical parameters $\{M,J=Ma\}$ in (\ref{Eq2}) are respectively the mass and angular momentum \cite{Noteaa} of the spinning 
Kerr black-hole spacetime.
The roots of the metric function $\Delta(r)$ determine the characteristic spin-dependent horizon radii
\begin{equation}\label{Eq3}
r_{\pm}=M\pm\sqrt{M^2-a^2}\  
\end{equation}
of the black hole. 

Interestingly, the spinning and curved black-hole spacetime (\ref{Eq1}) is characterized by a non-trivial (non-zero) 
spatially-dependent invariant  
\begin{equation}\label{Eq4}
{\cal G}\equiv R_{\mu\nu\rho\sigma}R^{\mu\nu\rho\sigma}-4R_{\mu\nu}R^{\mu\nu}+R^2\  ,
\end{equation}
known as the Gauss-Bonnet curvature invariant. 
From Eq. (\ref{Eq4}) one finds that the Gauss-Bonnet curvature invariant of the 
spinning Kerr black-hole spacetime (\ref{Eq1}) is given by the two-dimensional functional expression \cite{Donnw1,Donnw2}
\begin{equation}\label{Eq5}
{\cal G}_{\text{Kerr}}(r,\theta)={{48M^2}\over{(r^2+a^2\cos^2\theta)^6}}
\cdot\big({r^6-15r^4a^2\cos^2\theta+15r^2a^4\cos^4\theta-a^6\cos^6\theta}\big)\  .
\end{equation}
For a given value of the black-hole rotation parameter $a$, the expression (\ref{Eq5}) 
is a two-dimensional function of the radial coordinate
\begin{equation}\label{Eq6}
r\in[M+\sqrt{M^2-a^2},\infty]\
\end{equation}
and the polar coordinate
\begin{equation}\label{Eq7}
\theta\in[0,\pi]\ \Rightarrow \ \cos^2\theta\in[0,1]\  .
\end{equation}

The Gauss-Bonnet curvature invariant (\ref{Eq4}) has recently attracted the attention of many physicists and mathematicians. 
In particular, it has been revealed (see \cite{Donnw1,Donnw2,Dons,ChunHer,Hodca,Herkn} and references therein) that the influential 
no-hair conjecture \cite{NHC,JDB} can be 
violated in composed Einstein-Gauss-Bonnet-scalar field theories whose action
\begin{equation}\label{Eq8}
S={1\over2}\int
d^4x\sqrt{-g}\Big[R-{1\over2}\nabla_{\alpha}\phi\nabla^{\alpha}\phi+f(\phi){\cal
G}\Big]\
\end{equation}
contains a direct (non-minimal) coupling of the scalar field to the spatially-dependent Gauss-Bonnet invariant. 
Interestingly, it has been proved \cite{Dons,Donnw1,Donnw2,ChunHer,Hodca,Herkn} that the boundary between bald spinning 
Kerr black holes and hairy black-hole configurations in the Einstein-Gauss-Bonnet field theory (\ref{Eq8}) is marked 
by the presence of `cloudy' Kerr black holes that support linearized scalar fields with a non-minimal coupling 
to the Gauss-Bonnet invariant (\ref{Eq5}) of the curved spacetime (\ref{Eq1}).

Since the Gauss-Bonnet curvature invariant ${\cal G}_{\text{Kerr}}(r,\cos\theta;a/M)$ plays 
the role of an effective {\it spatially-dependent} mass term in the Klein-Gordon wave equation of the 
Einstein-Gauss-Bonnet-scalar field theory (\ref{Eq8}) \cite{Donnw1,Donnw2,Dons,ChunHer,Hodca,Herkn}, 
it is of physical interest to explore its highly non-trivial two-dimensional spatial functional behavior 
in the exterior region of the black-hole spacetime.

The main goal of the present paper is to explore, using {\it analytical} techniques, 
the physical and mathematical properties of the two-dimensional 
Gauss-Bonnet curvature invariant ${\cal G}_{\text{Kerr}}(r,\cos\theta;a/M)$ of astrophysically realistic 
spinning Kerr black holes. 
In particular, below we shall derive remarkably compact analytical formulas for the 
characteristic spin-dependent global extremum points of the Gauss-Bonnet invariant in the 
exterior region (\ref{Eq6}) of the rotating Kerr spacetime (\ref{Eq1}). 
Interestingly, we shall reveal the existence of two critical black-hole rotation parameters, 
$(a/M)^{-}_{\text{crit}}=1/2$ and $(a/M)^{+}_{\text{crit}}\simeq0.7818$ 
[see below the exact analytically derived dimensionless expression (\ref{Eq46})], which determine three qualitatively 
different global functional behaviors of the spin-dependent Kerr 
Gauss-Bonnet curvature invariant (\ref{Eq5}).
 
\section{Spatial behavior of the external two-dimensional Gauss-Bonnet 
curvature invariant ${\cal G}_{\text{Kerr}}(r,\theta)$ within its domain of existence}

In the present section we shall 
search for local extremum points and saddle points of the two-dimensional Kerr 
Gauss-Bonnet invariant (\ref{Eq5}) {\it within} the 
physically allowed domain [see Eqs. (\ref{Eq6}) and (\ref{Eq7})]
\begin{equation}\label{Eq9}
r\in(M+\sqrt{M^2-a^2},\infty)\ \ \ \ \text{with}\ \ \ \ \cos^2\theta\in(0,1)\  .
\end{equation}
The condition
\begin{equation}\label{Eq10}
{{\partial {\cal G}_{\text{Kerr}}(r,\theta)}\over{\partial r}}=0\
\end{equation}
yields the effectively cubic equation
\begin{equation}\label{Eq11}
r^6-21r^4a^2\cos^2\theta+35r^2a^4\cos^4\theta-7a^6\cos^6\theta=0\  ,
\end{equation}
whereas the condition
\begin{equation}\label{Eq12}
{{\partial {\cal G}_{\text{Kerr}}(r,\theta)}\over{\partial\theta}}=0\
\end{equation}
yields the effectively cubic equation \cite{Notecos0}
\begin{equation}\label{Eq13}
7r^6-35r^4a^2\cos^2\theta+21r^2a^4\cos^4\theta-a^6\cos^6\theta=0\  .
\end{equation}

From Eqs. (\ref{Eq11}) and (\ref{Eq13}) one finds the extremum condition 
\begin{equation}\label{Eq14}
7r^4-14r^2a^2\cos^2\theta+3a^4\cos^4\theta=0\  .
\end{equation}
The solution of (\ref{Eq14}) that can respect the condition (\ref{Eq9}) is 
given by the dimensionless relation
\begin{equation}\label{Eq15}
\cos\theta=\pm\sqrt{{{7-2\sqrt{7}}\over{3}}}\cdot {{r}\over{a}}\  .
\end{equation}
Substituting (\ref{Eq15}) into the relation
\begin{equation}\label{Eq16}
d\equiv {{\partial^2 {\cal G}_{\text{Kerr}}(r,\theta)}\over{\partial r^2}}\cdot
{{\partial^2 {\cal G}_{\text{Kerr}}(r,\theta)}\over{\partial\theta^2}}-
\Big[{{\partial^2 {\cal G}_{\text{Kerr}}(r,\theta)}\over{\partial r\partial\theta}}\Big]^2\  ,
\end{equation}
one finds \cite{Notedn}
\begin{equation}\label{Eq17}
d<0\  ,
\end{equation}
which implies that the two-dimensional Gauss-Bonnet curvature invariant (\ref{Eq5}) has no local extremum points within 
the domain (\ref{Eq9}).

\section{Functional behavior of the external Gauss-Bonnet curvature invariant ${\cal G}_{\text{Kerr}}(r,\theta)$ 
along its angular boundaries}

In the present section we shall analyze the functional behavior of the Kerr Gauss-Bonnet curvature invariant (\ref{Eq5}) 
along the angular boundaries [see Eq. (\ref{Eq7})] of the black-hole spacetime (\ref{Eq1}).  

\subsection{Analysis of the Gauss-Bonnet curvature invariant along the equatorial boundary $\cos^2\theta=0$}

Substituting the equatorial boundary relation 
\begin{equation}\label{Eq18}
\cos\theta=0\
\end{equation}
into Eq. (\ref{Eq5}), one obtains the remarkably simple functional expression
\begin{equation}\label{Eq19}
{\cal G}_{\text{Kerr}}(r,\cos\theta=0)={{48M^2}\over{r^6}}\  ,
\end{equation}
which is a monotonically decreasing function of the radial coordinate $r$ whose maximum value is obtained at 
the black-hole outer horizon:
\begin{equation}\label{Eq20}
{\cal G}_{\text{Kerr}}(r=M+\sqrt{M^2-a^2},\cos\theta=0)={{48M^2}\over{(M+\sqrt{M^2-a^2})^6}}\  .
\end{equation}

\subsection{Analysis of the Gauss-Bonnet curvature invariant along the polar boundary $\cos^2\theta=1$}

Substituting the polar boundary relation 
\begin{equation}\label{Eq21}
\cos^2\theta=1\
\end{equation}
into Eq. (\ref{Eq5}), one obtains the spin-dependent curvature expression
\begin{equation}\label{Eq22}
{\cal G}_{\text{Kerr}}(r,\cos^2\theta=1)={{48M^2}\over{(r^2+a^2)^6}}
\cdot(r^2-a^2)(r^4-14a^2r^2+a^4)\  .
\end{equation}
Interestingly, one finds that the radial expression (\ref{Eq22}) has a non-trivial radial functional behavior. 
In particular, the Gauss-Bonnet function (\ref{Eq22}) has three radial extremum points which 
are determined by the effectively cubic equation
\begin{equation}\label{Eq23}
r^6-21a^2\cdot r^4+35a^4\cdot r^2-7a^6=0\  .
\end{equation}

Defining the dimensionless variable 
\begin{equation}\label{Eq24}
x\equiv \Big({{r}\over{a}}\Big)^2\  ,
\end{equation}
the characteristic equation (\ref{Eq23}) can be expressed in the form
\begin{equation}\label{Eq25}
7-35x+21x^2-x^3=0\  .
\end{equation}
The cubic radial equation (\ref{Eq25}) can be solved analytically to yield the spin-dependent radial  
extremum points of the Gauss-Bonnet invariant (\ref{Eq22}). 
In particular, one finds that the Gauss-Bonnet function (\ref{Eq22}) has one local radial minimum point and one 
local radial maximum point that, in principle, can satisfy the radial requirement (\ref{Eq6}) \cite{Notex3}: 
\begin{equation}\label{Eq26}
x_{\text{min}}=7+4\sqrt{7}\sin\Big[{1\over3}\arctan\Big({{1}\over{3\sqrt{3}}}\Big)\Big]-
4\sqrt{{{7}\over{3}}}\cos\Big[{1\over3}\arctan\Big({{1}\over{3\sqrt{3}}}\Big)\Big]
\end{equation}
and
\begin{equation}\label{Eq27}
x_{\text{max}}=7+8\sqrt{{{7}\over{3}}}\cos\Big[{1\over3}\arctan\Big({{1}\over{3\sqrt{3}}}\Big)\Big]
\end{equation}

Taking cognizance of Eqs. (\ref{Eq6}), (\ref{Eq24}), (\ref{Eq26}), and (\ref{Eq27}) one finds that, 
depending on the magnitude of the dimensionless black-hole rotation parameter $a/M$, 
the Gauss-Bonnet invariant (\ref{Eq22}) 
has three qualitatively different radial functional behaviors:
\newline
{Case I:} From Eqs. (\ref{Eq6}), (\ref{Eq24}), and (\ref{Eq27}) one finds that, in the dimensionless spin regime \cite{Notecranmax}
\begin{equation}\label{Eq28}
{{a}\over{M}}<{{2\sqrt{x_{\text{max}}}}\over{1+x_{\text{max}}}}
\  ,
\end{equation}
the expression (\ref{Eq22}) is a monotonically decreasing function in the external radial region (\ref{Eq6}) of the black-hole 
spacetime (\ref{Eq1}), whose maximum value 
\begin{equation}\label{Eq29}
{\cal G}_{\text{Kerr}}(r=M+\sqrt{M^2-a^2},\cos^2\theta=1)=
{{3(M^2-4a^2)[(M+\sqrt{M^2-a^2})^2-a^2]}\over{M^4(M+\sqrt{M^2-a^2})^4}}\
\end{equation}
is located on the black-hole outer horizon. 
Note that the curvature value of the Gauss-Bonnet invariant is larger at the maximum point (\ref{Eq20}) 
than at the maximum point (\ref{Eq29}). 
\newline
{Case II:} From Eqs. (\ref{Eq6}), (\ref{Eq24}), (\ref{Eq26}), and (\ref{Eq27}) one finds that, in the dimensionless spin regime \cite{Notecranmin}
\begin{equation}\label{Eq30}
0.4338\simeq{{2\sqrt{x_{\text{max}}}}\over{1+x_{\text{max}}}}\leq{{a}\over{M}}<
{{2\sqrt{x_{\text{min}}}}\over{1+x_{\text{min}}}}\simeq0.9749\  ,
\end{equation}
the Gauss-Bonnet invariant ${\cal G}_{\text{Kerr}}(r,\cos^2\theta=1)$ has a local maximum radial point at 
\begin{equation}\label{Eq31}
r_{\text{max}}=\sqrt{7+8\sqrt{{{7}\over{3}}}\cos\Big[{1\over3}\arctan\Big({{1}\over{3\sqrt{3}}}\Big)\Big]}\cdot a
\geq r_+\ 
\end{equation}
with
\begin{eqnarray}\label{Eq32}
{\cal G}_{\text{Kerr}}(r=r_{\text{max}},\cos^2\theta=1)&=&
{{48M^2}\over{a^6}}
\cdot{{(x_{\text{max}}-1)(x^2_{\text{max}}-14x_{\text{max}}+1)}\over{(x_{\text{max}}+1)^6}}
\  .
\end{eqnarray} 
One finds that, in the dimensionless spin regime (\ref{Eq30}), the maximum point (\ref{Eq32}) 
has a Gauss-Bonnet curvature value which is smaller than the corresponding curvature at the 
maximum point (\ref{Eq20}). 
\newline
{Case III:} From Eqs. (\ref{Eq6}), (\ref{Eq24}), and (\ref{Eq26}) 
one finds that, for highly spinning Kerr black holes in the dimensionless regime
\begin{equation}\label{Eq33}
{{a}\over{M}}\geq{{2\sqrt{x_{\text{min}}}}\over{1+x_{\text{min}}}}\simeq0.9749\  ,
\end{equation}
the Gauss-Bonnet invariant ${\cal G}_{\text{Kerr}}(r,\cos^2\theta=1)$ has a local minimum radial point at
\begin{eqnarray}\label{Eq34}
r_{\text{min}}&=&\sqrt{7+4\sqrt{7}\sin\Big[{1\over3}\arctan\Big({{1}\over{3\sqrt{3}}}\Big)\Big]-
4\sqrt{{{7}\over{3}}}\cos\Big[{1\over3}\arctan\Big({{1}\over{3\sqrt{3}}}\Big)\Big]}\cdot a 
\geq r_+\
\end{eqnarray}
with
\begin{eqnarray}\label{Eq35}
{\cal G}_{\text{Kerr}}(r=r_{\text{min}},\cos^2\theta=1)&=&
{{48M^2}\over{a^6}}
\cdot{{(x_{\text{min}}-1)(x^2_{\text{min}}-14x_{\text{min}}+1)}\over{(x_{\text{min}}+1)^6}}\nonumber \\
&\simeq&
-1.7581\cdot{{M^2}\over{a^6}}\  ,
\end{eqnarray}
and a local maximum radial point which is characterized by the properties (\ref{Eq31}) and (\ref{Eq32}). 
Interestingly, one finds that, in the dimensionless spin regime (\ref{Eq33}), the maximum point (\ref{Eq32}) 
has a curvature value which is smaller than the corresponding Gauss-Bonnet curvature at the 
maximum point (\ref{Eq20}). 

\section{Functional behavior of the Gauss-Bonnet curvature invariant ${\cal G}_{\text{Kerr}}(r,\theta)$ 
along its radial boundaries}

In the present section we shall analyze the functional behavior of the Gauss-Bonnet invariant (\ref{Eq5}) along the radial 
boundaries [see Eq. (\ref{Eq6})] of the external black-hole spacetime (\ref{Eq1}). 
We first point out that asymptotically flat Kerr black-hole spacetimes are characterized 
by the trivial asymptotic functional behavior 
\begin{equation}\label{Eq36}
{\cal G}_{\text{Kerr}}(r\to\infty,\cos\theta)\to0^+\  .  
\end{equation}

Substituting the spin-dependent horizon boundary relation 
\begin{equation}\label{Eq37}
r=r_+=M+\sqrt{M^2-a^2}\  
\end{equation}
into Eq. (\ref{Eq5}), one obtains the functional expression
\begin{eqnarray}\label{Eq38}
&{\cal G}_{\text{Kerr}}(r=r_+,\cos\theta)= \nonumber \\ &
{{48M^2[(M+\sqrt{M^2-a^2})^2-a^2\cos^2\theta][(M+\sqrt{M^2-a^2})^4-14a^2\cos^2\theta (M+\sqrt{M^2-a^2})^2+a^4\cos^4\theta]}\over{[(M+\sqrt{M^2-a^2})^2+a^2\cos^2\theta]^6}}\  . 
\end{eqnarray}
The Gauss-Bonnet curvature invariant (\ref{Eq38}) on the outer horizon of the spinning Kerr black hole is 
characterized by the relations [see Eqs. (\ref{Eq20}) and (\ref{Eq29})]
\begin{equation}\label{Eq39}
{\cal G}_{\text{Kerr}}(r=r_+,\cos^2\theta=0)=
{{48M^2}\over{(M+\sqrt{M^2-a^2})^6}}\
\end{equation}
and
\begin{equation}\label{Eq40}
{\cal G}_{\text{Kerr}}(r=r_+,\cos^2\theta=1)=
{{3(M^2-4a^2)[(M+\sqrt{M^2-a^2})^2-a^2]}\over{M^4(M+\sqrt{M^2-a^2})^4}}\  .
\end{equation}

Inspection of the (rather cumbersome) curvature expression (\ref{Eq38}) reveals that, as a function of the polar 
variable $\cos^2\theta$, it has three extremum angular points. 
In particular, from Eq. (\ref{Eq38}) one obtains the effectively cubic equation
\begin{equation}\label{Eq41}
7r^6_+-35r^4_+a^2\cos^2\theta+21r^2_+a^4\cos^4\theta-a^6\cos^6\theta=0\
\end{equation}
for the locations of the spin-dependent polar extremum points of the Gauss-Bonnet curvature invariant (\ref{Eq38}) 
along the black-hole horizon. 

Defining the dimensionless variable 
\begin{equation}\label{Eq42}
x\equiv \Big[{{a}\over{r_+(a/M)}}\Big]^2\cos^2\theta\  ,
\end{equation}
one finds that Eq. (\ref{Eq41}) can be written in the form
\begin{equation}\label{Eq43}
7-35x+21x^2-x^3=0\  .
\end{equation}
The cubic polar equation (\ref{Eq43}) can be solved analytically to yield the spin-dependent 
extremum angular points of the Gauss-Bonnet curvature invariant (\ref{Eq38}) along the polar angular direction of the black-hole horizon. 
In particular, one finds that 
the only solution of (\ref{Eq43}) that can respect the angular condition (\ref{Eq7}) is a minimum point of the curvature 
function (\ref{Eq38}) which is given by the dimensionless relation \cite{Notex2x3,Notexap}
\begin{equation}\label{Eq44}
x_{\text{min}}=7-4\sqrt{7}\sin\Big[{1\over3}\arctan\Big({{1}\over{3\sqrt{3}}}\Big)\Big]-
4\sqrt{{{7}\over{3}}}\cos\Big[{1\over3}\arctan\Big({{1}\over{3\sqrt{3}}}\Big)\Big]
\  .
\end{equation}

Taking cognizance of Eqs. (\ref{Eq7}), (\ref{Eq42}), and (\ref{Eq44}) one finds that, for spinning 
Kerr black holes in the dimensionless sub-critical regime \cite{Notecran}
\begin{equation}\label{Eq45}
{{a}\over{M}}<\Big({{a}\over{M}}\Big)_{\text{crit}}=
\sqrt{{{7+\sqrt{7}\cos\Big[{1\over3}\arctan\big(3\sqrt{3}\big)\Big]-
\sqrt{21}\sin\Big[{1\over3}\arctan\big(3\sqrt{3}\big)\Big]}\over{12}}}\  ,
\end{equation}
the extremum points of the curvature function (\ref{Eq38}) [as obtained from the cubic equation (\ref{Eq43})] 
are characterized by the non-physical relation $\cos^2\theta>1$ [see Eq. (\ref{Eq7})], in which 
case the Gauss-Bonnet curvature invariant (\ref{Eq38}) monotonically decreases 
in the polar angular regime (\ref{Eq7}) from the value (\ref{Eq39}) to the value (\ref{Eq40}). 

It is important to point out that, in the dimensionless spin regime $a/M\geq{1\over 2}$, the minimum point (\ref{Eq40}) at 
the poles of the black-hole surface is characterized by a non-positive value of the Gauss-Bonnet curvature invariant 
which is smaller than the asymptotic limit (\ref{Eq36}).

Intriguingly, one finds that, for rapidly spinning Kerr black holes 
in the complementary super-critical regime
\begin{equation}\label{Eq46}
{{a}\over{M}}\geq\Big({{a}\over{M}}\Big)_{\text{crit}}=
\sqrt{{{7+\sqrt{7}\cos\Big[{1\over3}\arctan\big(3\sqrt{3}\big)\Big]-
\sqrt{21}\sin\Big[{1\over3}\arctan\big(3\sqrt{3}\big)\Big]}\over{12}}}
\  ,
\end{equation}
one of the extremum angular points [the polar minimum point (\ref{Eq44})] of the curvature function (\ref{Eq38}) 
lies within the physically allowed angular region (\ref{Eq7}), in which case the Gauss-Bonnet curvature invariant (\ref{Eq38}) 
has a non-trivial (non-monotonic) angular functional behavior. 
In particular, one obtains from Eqs. (\ref{Eq3}), (\ref{Eq42}), and (\ref{Eq44}) the polar minimum point 
\begin{eqnarray}\label{Eq47}
(\cos^2\theta)_{\text{min}}= 
\big({{r_+}\over{a}}\big)^2\cdot
\Big\{7-4\sqrt{7}\sin\Big[{1\over3}\arctan\Big({{1}\over{3\sqrt{3}}}\Big)\Big]-
4\sqrt{{{7}\over{3}}}\cos\Big[{1\over3}\arctan\Big({{1}\over{3\sqrt{3}}}\Big)\Big]\Big\}
\  .
\end{eqnarray}
It is interesting to point out that the analytically derived formula (\ref{Eq47}) implies that the 
polar angle $\theta_{\text{min}}=\theta_{\text{min}}(a/M)$ (with $\theta_{\text{min}}\leq90^{\circ}$), which characterizes 
the minimum angular point of the Gauss-Bonnet curvature invariant (\ref{Eq38}), is 
a monotonically increasing function of the dimensionless black-hole rotation parameter $a/M$.  

Taking cognizance of Eqs. (\ref{Eq38}), and (\ref{Eq47}), one obtains the functional expression
\begin{eqnarray}\label{Eq48}
{\cal G}_{\text{Kerr}}[r=r_+(a/M),(\cos^2\theta)_{\text{min}}]=
-{{M^2}\over{r^6_+}}\cdot
{{57+28\sqrt{21}\cos\Big[{1\over3}\arctan\Big({{1}\over{3\sqrt{3}}}\Big)\Big]}\over{8}}
\  
\end{eqnarray}
for the spin-dependent minimal value of the Gauss-Bonnet curvature invariant (\ref{Eq38}) that characterizes the spinning Kerr 
spacetime (\ref{Eq1}) along the black-hole horizon. 
Interestingly, one finds that the absolute value of the analytically derived functional relation (\ref{Eq48}) is 
a monotonically increasing function of the dimensionless black-hole rotation parameter $a/M$. 

Taking cognizance of Eqs. (\ref{Eq35}) and (\ref{Eq48}), 
one finds that for rapidly-rotating Kerr black holes in the dimensionless spin 
regime (\ref{Eq46}), the value (\ref{Eq48}) of the 
curvature invariant ${\cal G}_{\text{Kerr}}[r=r_+(a/M),(\cos^2\theta)_{\text{min}}]$ 
is smaller (that is, more negative) 
than the curvature value ${\cal G}_{\text{Kerr}}(r=r_{\text{min}},\cos^2\theta=1)$ given by Eq. (\ref{Eq35}). 

In table \ref{Table1} we present, using the analytically derived formulas (\ref{Eq47}) and (\ref{Eq48}), the angular values of the 
polar minimum points $(\cos^2\theta)_{\text{min}}(a/M)$ and the corresponding values of the Gauss-Bonnet curvature 
invariant ${\cal G}_{\text{Kerr}}[r=r_+(a/,M),(\cos^2\theta)_{\text{min}}]$ for various values of the dimensionless black-hole rotation parameter $a/M$ in the super-critical regime (\ref{Eq46}). 

\begin{table}[htbp]
\centering
\begin{tabular}{|c|c|c|}
\hline $a/M$ & \ $(\cos^2\theta)_{\text{min}}$\ \ & \
$M^4{\cal G}_{\text{Kerr}}[r=r_+(a/M),(\cos^2\theta)_{\text{min}}]$\ \  \\
\hline \ $0.80$\ \ \ &\ \ 0.9277\ \ \ &\ \ -1.3788\ \ \\
\hline \ $0.85$\ \ \ &\ \ 0.7482\ \ \ &\ \ -1.8262\ \ \\
\hline \ $0.90$\ \ \ &\ \ 0.5903\ \ \ &\ \ -2.6393\ \ \\
\hline \ $0.95$\ \ \ &\ \ 0.4425\ \ \ &\ \ -4.5301\ \ \\
\hline \ $0.975$\ \ \ &\ \ 0.3644\ \ \ &\ \ -6.9398\ \ \\
\hline \ $0.999$\ \ \ &\ \ 0.2536\ \ \ &\ \ -17.7924\ \ \\
\hline \ $1.000$\ \ \ &\ \ 0.2319\ \ \ &\ \ -23.1318\ \ \\
\hline
\end{tabular}
\caption{The Gauss-Bonnet curvature invariant of spinning Kerr black holes. 
We present, for various super-critical values of the dimensionless black-hole 
rotation parameter $a/M$ [see Eq. (\ref{Eq46})], the values of $(\cos^2\theta)_{\text{min}}(a/M)$ which characterize the 
minimum angular points of the Gauss-Bonnet invariant (\ref{Eq38}) along the black-hole horizon as obtained 
from the analytically derived formula (\ref{Eq47}). 
We also present the corresponding spin-dependent values of the dimensionless Gauss-Bonnet curvature 
invariant ${\cal G}_{\text{Kerr}}[r=r_+(a/M),(\cos^2\theta)_{\text{min}}]$ as obtained 
from the analytically derived formula (\ref{Eq48}).}\label{Table1}
\end{table}

\section{Kerr black holes supporting infinitesimally thin massive scalar rings}

In the present section we shall reveal the fact that spinning Kerr black holes can support thin matter rings which 
are made of {\it massive} scalar fields with a non-minimal coupling to the Gauss-Bonnet invariant (\ref{Eq5}) 
of the curved spacetime. As we shall now show, this intriguing physical observation is a direct outcome of our 
analytically derived results.

The composed Einstein-Gauss-Bonnet-nonminimally-coupled-massive-scalar field 
theory is characterized by the action \cite{Donnw2}
\begin{equation}\label{Eq49}
S=\int
d^4x\sqrt{-g}\Big[{1\over4}R-{1\over2}\nabla_{\alpha}\phi\nabla^{\alpha}\phi
-{1\over2}\mu^2\phi^2+f(\phi){\cal G}\Big]\  .
\end{equation}
Here the physical parameter $\mu$ is the mass of the non-minimally coupled scalar field \cite{Notemuu}. 
It has been proved \cite{Dons,Donnw1,Donnw2,ChunHer,Hodca,Herkn} that the boundary between bald spinning 
Kerr black holes and hairy (scalarized) black-hole configurations in the Einstein-Gauss-Bonnet-scalar field 
theory (\ref{Eq49}) is marked by the presence of marginally stable Kerr black holes (`cloudy' Kerr black holes) 
that support linearized configurations of the non-minimally coupled massive scalar fields. 
Intriguingly, as we shall now show, the supported spatially regular external field configurations owe their existence to the 
non-minimal direct coupling term $f(\phi){\cal G}$ 
between the massive scalar field $\phi$ and the Gauss-Bonnet invariant (\ref{Eq5}) of the 
curved spacetime [see the action (\ref{Eq49})].

The action (\ref{Eq49}) yields the Klein-Gordon differential equation \cite{Donnw2}
\begin{equation}\label{Eq50}
\nabla^\nu\nabla_{\nu}\phi=\mu^2_{\text{eff}}\phi\
\end{equation}
for the non-minimally coupled massive scalar field configurations, where the spatially-dependent 
effective mass term in Eq. (\ref{Eq50}),
\begin{equation}\label{Eq51}
\mu^2_{\text{eff}}(r,\theta;M,a)=\mu^2-\eta\cdot{\cal G}_{\text{Kerr}}(r,\theta)\  ,
\end{equation}
reflects the direct massive-scalar-field-Kerr-Gauss-Bonnet coupling in the 
composed Einstein-Gauss-Bonnet-nonminimally-coupled-massive-scalar field theory (\ref{Eq49}). 
Here the physical parameter $\eta$ \cite{Noteetaa}, which appears in the weak-field expansion \cite{Donnw2}
\begin{equation}\label{Eq52}
f(\phi)={1\over2}\eta\phi^2\ 
\end{equation}
of the scalar coupling function, controls the strength of the direct (non-minimal) coupling 
between the massive scalar field configurations 
and the Gauss-Bonnet curvature invariant (\ref{Eq5}). 

Interestingly, taking cognizance of Eq. (\ref{Eq5}), one finds that, depending on the relative magnitudes of the physical 
parameters $\eta$ and $\mu$ of the composed Einstein-Gauss-Bonnet-nonminimally-coupled-massive-scalar field 
theory (\ref{Eq49}), the spatially-dependent effective mass term (\ref{Eq51}) of the 
non-minimally coupled scalar field may become negative in the vicinity of the outer horizon of the central 
supporting Kerr black hole. 

The presence of an effective binding ({\it negative}) potential well outside the 
outer horizon of the supporting black hole provides a necessary condition for the existence of spatially regular 
bound-state scalar configurations (scalar clouds) that are supported in the asymptotically flat 
curved black-hole spacetime (\ref{Eq1}) \cite{Dons,Donnw1,Donnw2,ChunHer,Hodca,Herkn}. 

In particular, for a given mass $\mu$ of the non-minimally coupled scalar field in the dimensionless 
large-mass regime 
[or equivalently, in the dimensionless large-coupling $\eta/M^2\gg1$ regime, see Eq. (\ref{Eq55}) below]
\begin{equation}\label{Eq53}
M\mu\gg1\  ,
\end{equation}
the {\it onset} of the spontaneous scalarization phenomena is marked by the critical relation \cite{Hodca,Herkn,Hodjp}
\begin{equation}\label{Eq54}
\text{min}\{\mu^2_{\text{eff}}(r,\theta;M,a)\}\to 0^{-}\
\end{equation}
of the effective mass term in the characteristic Klein-Gordon differential equation (\ref{Eq50}). 

Taking cognizance of Eqs. (\ref{Eq20}), (\ref{Eq51}), and (\ref{Eq54}) one finds that, in the 
large-mass regime (\ref{Eq53}) with
\begin{equation}\label{Eq55}
{{48M^2}\over{(M+\sqrt{M^2-a^2})^6}}\cdot{{\eta}\over{\mu^2}}\to 1^{+}\  ,
\end{equation}
the effective mass term (\ref{Eq51}) of the composed Kerr-black-hole-nonminimally-coupled-massive-scalar-field system 
becomes {\it negative} (attractive) in the narrow equatorial 
{\it ring} which is characterized by the relations $r\to r_+(a,M)$ with $\theta\to\pi/2$. 
This physically interesting fact implies that, in the dimensionless large-mass regime (\ref{Eq53}), 
spinning Kerr black holes can support infinitesimally thin non-minimally 
coupled massive scalar configurations (scalar rings) which are characterized by the dimensionless 
ratio (\ref{Eq55}) and are located on the equator of the black-hole surface. 

\section{Summary and discussion}

Motivated by the recent growing interest in composed Einstein-Gauss-Bonnet field theories and the presence 
of cloudy Kerr black holes \cite{Notekb} that support external matter configurations with a direct non-minimal coupling 
to the Gauss-Bonnet invariant (see \cite{Donnw1,Donnw2,Dons,ChunHer,Hodca,Herkn} and references therein), 
we have explored the physical and mathematical properties of the two-dimensional 
Gauss-Bonnet curvature invariant ${\cal G}_{\text{Kerr}}(r,\cos\theta;a/M)$ of astrophysically realistic 
spinning Kerr black holes. 

The main {\it analytical} results derived in this paper are as follows:

(1) We have proved that the global maximum point \cite{Notering} of the Kerr Gauss-Bonnet curvature invariant in the 
exterior region $r\in[r_+(a/M),\infty]$ of the black-hole spacetime is always 
(that is, for all Kerr black holes in the physically allowed regime $a/M\in[0,1]$) 
located on the black-hole horizon. 
In particular, taking cognizance of Eqs. (\ref{Eq20}), (\ref{Eq29}), and (\ref{Eq32}), 
one finds that the global maximum value 
\begin{equation}\label{Eq56}
\text{max}\Big\{M^4\cdot{\cal G}_{\text{Kerr}}(r\in[r_+,\infty],\cos^2\theta;a/M)\Big\}=
{{48M^6}\over{(M+\sqrt{M^2-a^2})^6}}\
\end{equation}
of the Gauss-Bonnet curvature invariant is 
located on the equator ($\theta_{\text{max}}=90^{\circ}$) of the black-hole surface. 

(2) The analytically derived formula (\ref{Eq56}) implies that the global maximum value 
of the Gauss-Bonnet curvature invariant is a monotonically increasing function 
of the dimensionless black-hole rotation parameter $a/M$. 
In particular, the expression (\ref{Eq56}) yields the dimensionless curvature value  
\begin{equation}\label{Eq57}
\text{max}\Big\{M^4\cdot{\cal G}_{\text{Kerr}}(r\in[r_+,\infty],\cos^2\theta;a/M=1)\Big\}=48\
\end{equation}
for the maximally-spinning extremal Kerr black hole. 
This is the largest curvature value that characterizes the spin-dependent external Gauss-Bonnet 
invariant (\ref{Eq5}) of rotating Kerr black-hole spacetimes. 

(3) Intriguingly, we have proved that the location of the global minimum point which characterizes 
the Gauss-Bonnet curvature invariant of spinning Kerr black holes has a 
non-trivial functional dependence on the black-hole rotation parameter. 
In particular, we have revealed the existence of two critical black-hole rotation parameters:
\begin{equation}\label{Eq58}
\Big({{a}\over{M}}\Big)^{-}_{\text{crit}}={1\over2}\
\end{equation}
and
\begin{equation}\label{Eq59}
\Big({{a}\over{M}}\Big)^{+}_{\text{crit}}=
\sqrt{{{7+\sqrt{7}\cos\Big[{1\over3}\arctan\big(3\sqrt{3}\big)\Big]-
\sqrt{21}\sin\Big[{1\over3}\arctan\big(3\sqrt{3}\big)\Big]}\over{12}}}
\  ,
\end{equation}
which mark the boundaries between three qualitatively different functional behaviors of the 
Gauss-Bonnet curvature invariant: 
\newline
(i) For Kerr black holes in the sub-critical regime $a/M<(a/M)^{-}_{\text{crit}}$, the 
Gauss-Bonnet curvature invariant attains its global minimum asymptotically 
at spatial infinity [see Eq. (\ref{Eq36})]. 
\newline
(ii) Kerr lack holes in the intermediate regime $(a/M)^{-}_{\text{crit}}\leq a/M\leq(a/M)^{+}_{\text{crit}}$ 
are characterized by Gauss-Bonnet curvature invariants whose 
global minima are located at the black-hole poles
\newline
\begin{eqnarray}\label{Eq60}
(\cos^2\theta)_{\text{min}}=1\ \ \ \ \ \text{for}\ \ \ \ \ (a/M)^{-}_{\text{crit}}\leq a/M\leq(a/M)^{+}_{\text{crit}}\  .
\end{eqnarray}
\newline
(iii) Rapidly-spinning Kerr black holes in the 
super-critical regime
$a/M>(a/M)^{+}_{\text{crit}}$ 
are characterized by Gauss-Bonnet curvature invariants with non-monotonic functional behaviors along 
the polar angular direction of the black-hole surface. In particular, 
the spin-dependent global minima of the Gauss-Bonnet curvature invariants of these rapidly-rotating Kerr 
black holes are determined by the analytically derived dimensionless scaling relation [see Eq. (\ref{Eq47})]
\begin{eqnarray}\label{Eq61}
\Big({{a}\over{r_+}}\Big)^2\cdot(\cos^2\theta)_{\text{min}}&=&
7-4\sqrt{7}\sin\Big[{1\over3}\arctan\Big({{1}\over{3\sqrt{3}}}\Big)\Big]-
4\sqrt{{{7}\over{3}}}\cos\Big[{1\over3}\arctan\Big({{1}\over{3\sqrt{3}}}\Big)\Big]\nonumber \\ &&
\text{for}\ \ \ \ a/M\geq(a/M)^{+}_{\text{crit}}\  . 
\end{eqnarray}

(4) From the analytically derived formula (\ref{Eq61}) one learns that the polar angle $\theta_{\text{min}}$ 
(with $\theta_{\text{min}}\leq90^{\circ}$), which characterizes the minimum angular point of the Kerr Gauss-Bonnet 
curvature invariant in the super-critical regime $a/M\geq(a/M)^{+}_{\text{crit}}$, is 
a monotonically increasing function of the dimensionless black-hole spin parameter $a/M$. 
In particular, one finds from Eq. (\ref{Eq61}) the relation \cite{Noteth2}
\begin{equation}\label{Eq62}
\theta^{-}_{\text{min}}(a/M=1)\simeq61.212^{\circ}\ 
\end{equation}
for maximally-spinning (extremal) Kerr black holes. 
It is interesting to emphasize the fact that the 
value $\theta^{-}_{\text{min}}\simeq61.212^{\circ}$ is the {\it largest} polar angle (with $\theta_{\text{min}}\leq90^{\circ}$) that characterizes the spin-dependent global minimum points of the Gauss-Bonnet invariants 
of curved Kerr black-hole spacetimes.  

(5) Taking cognizance of Eqs. (\ref{Eq35}), (\ref{Eq36}), (\ref{Eq40}), and (\ref{Eq48}) one concludes that, 
in the exterior region (\ref{Eq6}) of the spinning Kerr black-hole spacetime, the 
Gauss-Bonnet curvature invariant is characterized by the global dimensionless minimum
\begin{eqnarray}\label{Eq63}
&{\text{min}\Big\{M^4\cdot{\cal G}_{\text{Kerr}}(r\in[r_+,\infty],\cos^2\theta;a/M)\Big\}}=
\nonumber\\ 
&
\begin{cases}
0^{+} & \ \ \ \text{for}\ \ \ \ \ \ a/M<(a/M)^{-}_{\text{crit}}\ \\
{{3(M^2-4a^2)[(M+\sqrt{M^2-a^2})^2-a^2]}\over{(M+\sqrt{M^2-a^2})^4}} & \ \ \ \text{for}\ \ \ \ \ 
\ (a/M)^{-}_{\text{crit}}\leq a/M\leq(a/M)^{+}_{\text{crit}}\ \\
-{{57+28\sqrt{21}\cos\big[{1\over3}\arctan\big({{1}\over{3\sqrt{3}}}\big)\big]}\over{8}}\cdot
{{M^6}\over{(M+\sqrt{M^2-a^2})^6}} & \ \ \ \text{for}\ \ \ \ \ \ a/M\geq(a/M)^{+}_{\text{crit}}\  . 
\end{cases}
\end{eqnarray}

(6) Interestingly, one learns from the analytically derived formula (\ref{Eq63}) that, in 
the dimensionless spin regime $a/M\geq(a/M)^{+}_{\text{crit}}$, the minimum value 
of the Gauss-Bonnet curvature invariant is a monotonically decreasing function 
of the dimensionless black-hole rotation parameter $a/M$. 
In particular, the expression (\ref{Eq63}) yields the dimensionless relation  
\begin{eqnarray}\label{Eq64}
\text{min}\Big\{M^4\cdot{\cal G}_{\text{Kerr}}(r\in[r_+,\infty],\cos^2\theta;a/M=1)\Big\}&=&
-{{57+28\sqrt{21}\cos\Big[{1\over3}\arctan\Big({{1}\over{3\sqrt{3}}}\Big)\Big]}\over{8}}
\
\end{eqnarray}
for the maximally-spinning extremal Kerr black hole. 
This is the most negative curvature value that characterizes the spin-dependent external Gauss-Bonnet 
invariant of rotating Kerr black-hole spacetimes. 


(7) We have proved that, in the large-mass regime (\ref{Eq53}) of 
the composed Einstein-Gauss-Bonnet-nonminimally-coupled-massive-scalar field theory (\ref{Eq49}), 
spinning Kerr black holes can support infinitesimally thin configurations of the non-minimally 
coupled massive scalar fields. These supported scalar clouds (massive scalar rings) 
are located on the equator of the black-hole surface and 
are characterized by the dimensionless large-mass (or equivalently, large-coupling) critical relation [see Eqs. (\ref{Eq3}) 
and (\ref{Eq55})] 
\begin{equation}\label{Eq65}
{{48M^2}\over{r^6_+}}\cdot{{\eta}\over{\mu^2}}\to 1^{+}\  .
\end{equation}

\bigskip
\noindent
{\bf ACKNOWLEDGMENTS}
\bigskip

This research is supported by the Carmel Science Foundation. I would
like to thank Yael Oren, Arbel M. Ongo, Ayelet B. Lata, and Alona B.
Tea for helpful discussions.



\begin{thebibliography}{99}
\bibitem{Notetwo} We use here the term `two-dimensional' to emphasize the fact that the 
Kerr black-hole spacetime is characterized by two conserved physical parameters: the asymptotically measured 
mass and angular momentum of the spinning spacetime.
\bibitem{Noteun} We use natural units in which $8\pi G=c=\hbar=1$.
\bibitem{Noteaa} We shall assume $a\geq0$ without loss of generality.
\bibitem{Notecos0} Note that in the present section we analyze the functional behavior of the two-dimensional 
Gauss-Bonnet invariant inside its domain of existence, in which case $\cos\theta\neq0$ [see Eq. (\ref{Eq9})].
\bibitem{Notedn} It is worth noting that the expression (\ref{Eq16}) is negative definite for the Kerr 
Gauss-Bonnet invariant, implying that the two-dimensional curvature function (\ref{Eq5}) has no local extremum points.
\bibitem{Notex3} Note that the third solution of the cubic equation (\ref{Eq25}), 
which corresponds to a local maximum radial point of the function (\ref{Eq22}), is smaller than 1. 
Thus, it cannot satisfy the radial relation (\ref{Eq6}) [see Eq. (\ref{Eq24})].
\bibitem{Notecranmax} Note that the critical rotation parameter in (\ref{Eq28}) is obtained by substituting 
the marginally allowed value $r=r_+=M+\sqrt{M^2-a^2}$ of the external radial coordinate 
into Eqs. (\ref{Eq24}) and (\ref{Eq27}).
\bibitem{Notecranmin} Note that the critical rotation parameters in (\ref{Eq30}) are obtained by substituting 
the marginally allowed value $r=r_+=M+\sqrt{M^2-a^2}$ of the external radial coordinate 
into Eqs. (\ref{Eq24}), (\ref{Eq26}), and (\ref{Eq27}).
\bibitem{Notex2x3} Note that the other two solutions of the cubic equation (\ref{Eq43}), 
which correspond to local maximum angular points of the curvature function (\ref{Eq38}), are larger than 1. 
Hence, they cannot satisfy the angular relation (\ref{Eq7}) [see Eq. (\ref{Eq42})].
\bibitem{Notexap} Since $x\leq1$ for the physically allowed values of $(\cos^2\theta)_{\text{min}}$ 
[see Eqs. (\ref{Eq3}), (\ref{Eq7}), and (\ref{Eq42})], 
one can approximate the cubic equation (\ref{Eq43}) by the simpler quadratic equation $7-35x+21x^2=0$, 
which yields $x^{\text{approx}}_{\text{min}}=(5-\sqrt{13})/6\simeq0.2324$ for the physically acceptable solution. 
Interestingly, one finds that this approximated value agrees with the exact analytically derived value (\ref{Eq44}) 
to within $\sim0.2\%$.
\bibitem{Notecran} Note that the critical rotation parameter (\ref{Eq45}) is obtained by substituting 
the marginally allowed angular value $\cos^2\theta=1$ 
into the analytically derived expression [see Eq. (\ref{Eq47}) below] 
for the polar location of the spin-dependent minimum angular 
point of the Gauss-Bonnet curvature invariant (\ref{Eq38}) along the black-hole horizon.
\bibitem{Notemuu} The scalar field mass parameter $\mu$ stands for $\mu/\hbar$ and 
therefore has the dimensions of $\text{length}^{-1}$.
\bibitem{Noteetaa} The non-trivial coupling parameter $\eta$ of the composed 
Einstein-Gauss-Bonnet-nonminimally-coupled-massive-scalar field theory (\ref{Eq49}) 
has the dimensions of length$^2$.
\bibitem{Notekb} It is worth noting again that the term `cloudy' black holes refers here to 
Kerr black holes that support linearized scalar fields. 
The composed Kerr-black-hole-linearized-scalar-field cloudy configurations mark 
the boundary between bald Kerr black holes and hairy spinning 
black-hole configurations of the Einstein-Gauss-Bonnet field theory (\ref{Eq8}).
\bibitem{Notering} Note that, due to the azimuthal symmetry of the Kerr black-hole spacetime (\ref{Eq1}), 
the $\{r,\theta\}$-dependent Gauss-Bonnet curvature invariant (\ref{Eq5}) is characterized by extremum rings with $\phi\in[0,2\pi]$.
\bibitem{Noteth2} Note that, due to the reflection symmetry of the Kerr spacetime, 
the second angular minimum point (with $\theta_{\text{min}}>90^{\circ}$) of the Kerr Gauss-Bonnet 
curvature invariant is given by the simple relation $\theta^{+}_{\text{min}}=180^{\circ}-\theta^{-}_{\text{min}}$.
\end{thebibliography}
\end{document}